\documentclass[12pt,article,amssymb,amsmath,amsfonts,preprintnumbers,nofootinbib]{revtex4}

\def\be{\begin{equation}}
\def\ee{\end{equation}}
\def\bea{\begin{eqnarray}}
\def\eea{\end{eqnarray}}

\newcommand\bra[1]{{\langle {#1}|}}
\newcommand\ket[1]{{|{#1}\rangle}}

\def\Qh{\hat{Q}}

\begin{document}
\begin{flushright}
\preprint{UPR-1173-T}
\end{flushright}

\title{4d-5d connection and the holomorphic anomaly}
\author{Peng Gao}\thanks{The author is supported in part by DOE grant DE-FG02-95ER40893.}
\vskip2truecm

\affiliation{David Rittenhouse Laboratory,University of Pennsylvania,\\Philadelphia, PA 19104, USA.}
\date{\today}

\begin{abstract}
{\bf Abstract} In this short note, we report a curious appearance of the recently discovered 4d-5d connection of extremal blackholes in the topological string B-model. The holomorphic anomaly equations in the Schr${\rm o}$dinger-Weil representation are written {\it formally} in terms of M2 charges. In the phase space the 4d-5d charges are related by a {\it non-linear} canonical transformation. The blackhole partition function factors into M2-anti-M2 contributions in leading approximation.
\end{abstract}\maketitle

\newpage

\section{Introduction}\label{s1}
The study of microscopic entropy \cite{Strominger:1996sh} of attractor blackholes \cite{Ferrara:1995ih} has recently been revived largely due to the discovery of a deep connection with the topological string partition function \cite{Ooguri:2004zv}. In this note, we report some interesting findings in the topological B-model holomorphic anomaly equation which bears resemblance to the 4d-5d connection of extremal blackholes \cite{Gaiotto:2005gf},\cite{Bena:2005ni} (see also \cite{Dijkgraaf:2006um}). 

First we briefly review some recent developments on the holomorphic anomaly equations. The holomorphic anomaly was discovered from contributions of degenerate (bubbling) Riemann surfaces to the partition function (\cite{Bershadsky:1993ta} also see \cite{Antoniadis:1993ze}). The BCOV holomorphic anomaly signals a base point dependence of the topological string partition function on the  moduli space.  Witten then showed this apparent paradox may be reconciled by viewing the partition function as the wave function of a background independent theory expressed in a base-point dependent basis \cite{Witten:1993ed}. 

This approach was further developed in \cite{Verlinde:2004ck} where the wave function was obtained by (first-)quantizing a seven dimensional abelian Chern-Simons theory whose phase space is $H^3(X,\mathbb{R})$ with $X$ the Calabi-Yau manifold. In a related development \cite{Gerasimov:2004yx}, it was shown the Kodaira-Spencer path integral can be obtained from a projector imposing the physical state condition and the overlap of wavefunctions between the holomorphic polarization and a decomposible nonlinear one. 
More recently \cite{Gunaydin:2006bz}, a new holomorphic polarization has been found where the holomorphic anomaly equations take the form of generalized heat equations as also first suggested in \cite{Witten:1993ed}. 

Furthermore, the authors of  \cite{Gunaydin:2006bz} gave the holomorphic wave function a representation theory interpretation when the moduli space is a symmetric special Kahler manifold ${\cal M}={\rm G/K}$. Such spaces arise as vector multiplet moduli spaces of the so-called magic ${\cal N}=2$ supergravity theories\footnote{One of these theories is known to be the compactification of type II string theory on the Enriques Calabi-Yau\cite{Ferrara:1995yx}.} each associated to a Jordan algebra $J$ \cite{Gunaydin:1983rk}. Over $J$ one can define a real vector space V equipped with a natural symplectic marking called the Freudenthal triple \cite{frt} in which the electric and magnetic charges take values \cite{Ferrara:1997uz}. This is the counterpart of the real polarization of $H^3(X,\mathbb{R})$ for type II string on a Calabi-Yau manifold. For details of G group action on V and the construction of the moduli space ${\cal M}$ from V, we refer the readers to \cite{Gunaydin:2006bz}, \cite{Gunaydin:2005mx}. For our purpose it suffices to recall that the linear coordinates $(p^I,q_I)=(p^0,p^i,q_0,q_i)$ under the decomposition 
\be
V=\mathbb{R}\oplus J\oplus J\oplus\mathbb{R}
\ee
are canonical variables on the phase space V. Upon quantization they satisfy the commutation rules
\be
[\hat{p}^I,\hat{q}_J]=Z\delta^I_J
\ee
and generate the Heisenberg group H.  Together with generators of G given in terms of $(\hat{p}^I,\hat{q}_J)$, they generate the Jacobi group ${\rm G}\ltimes{\rm H}$ of which the Hilbert space form the Schr${\rm\ddot o}$dinger-Weil representation.

\section{5D charges in phase space}\label{s3}

Following \cite{Gunaydin:2006bz}, the nontrivial holomorphic anomaly equation can be written as
\be\label{gnp}
Z\hat T_i+{1\over2}C_{ijk}\hat p^j\hat p^k-\hat p^0\hat q_i=0~,
\ee
which holds as an operator identity in the Schr${\rm \ddot o}$dinger-Weil representation. On coherent states, $\hat T_i$ acts as ${\partial\over\partial t^i}$ and we may view this equation as a (multi-time) Schr${\rm \ddot o}$dinger equation with Hamiltonian(s)
\be
{\cal H}_i=\hat p^0\hat q_i-{1\over2}C_{ijk}\hat p^j\hat p^k
\ee
The central generator $Z=i\hbar$ and we take $\hbar=1$ from now on unless otherwise specified. The topological string partition function is naturally interpreted as the wave function of a state in the Hilbert space.

 In view of the relation between 4d and 5d extremal blackhole charges \cite{Gaiotto:2005gf}, we define the following operator
\be
\hat Q_i=\hat p^0\hat q_i-{1\over2}C_{ijk}\hat p^j\hat p^k=\hat{q_{i}}^{5d} 
\ee 
which is the operator corresponding to the charge of M2-branes of the 5d black-hole(ring)! In other words, we have found
\be
{\cal H}_i=\hat Q_i
\ee 
with $Q_i$ the formally defined M2-charges. Next we show we can in fact interpret this new definition of operators as a canonical transformation in the phase space. 

Classically we look for the new conjugate variables which turn out to be given by the following pairs $(\tilde p^0,\tilde q_0)=(-q_0-{C_{ijk}p^ip^jp^k\over6p^0{}^2},p^0)$ and $(\tilde p^i,\tilde q_i)=(p^i,Q_i/p^0)$. In terms of the 5d charges $(p^i,q_i^{5d},J_L^3)$, the new pairs are naturally written as $({p^iq_i^{5d}\over p^0}-2J_L^3,p^0)$ and $(p^i,q_i^{5d}/p^0)$. In the M-theory 'lift', $p^0$ is  a nut charge, $q_0$ momentum along the M-theory circle, and $q_i^{5d},p^i$ the charges and dipole charges respectively. The particular combination ${p^iq_i^{5d}\over p^0}-2J_L^3$ is the contribution to angular momentum purely from M2 bound states. The symplectic form is preserved by the transformation
\bea
\omega&=&-d(q_0+{C_{ijk}p^ip^jp^k\over6p^0{}^2})\wedge dp_0+dp^i\wedge d(Q_i/p^0)\nonumber\\
&=&dp^0\wedge dq_0+dp^i\wedge dq_i~
\eea
which guarantees the consistency of quantization. Here we used the fact that $C_{ijk}$ is constant and symmetric in all its indices. In the notation of \cite{Gunaydin:2006bz}, the grading of $Q_i/p^0$ is $+2$ the same as that of $q_i$ and our redefinition of coordinates preserves the orbits of G.

It's easy to see the above canonical transformation is generated by 
\bea\label{HJ}
&&\tilde p^Id\tilde q_I-p^Idq_I=dS_{54}~,\nonumber\\
&& S_{54}=-p^0q_0-{C_{ijk}p^ip^jp^k\over3p^0}=-p^0q_0-2{N(p)\over p^0}~.
\eea
The second term of this generating function is nothing but twice the purely cubic tree-level prepotential $F(p)$!  As another property of these special ${\cal N}=2$ theories, the prepotentials are all invariant under Fourier transform with respect to all charges at once \cite{Pioline:2005vi}. 

Quantum mechanically, it's straightforward to check that the operators corresponding to the new charges satisfy the expected commutation relations
\bea
&&[\Qh_i,\Qh_j]=0,\quad [\Qh_i,\hat p^0]=0\nonumber\\
&&[\hat{p}^i, \hat{\tilde{q}}_j]={\rm i}\delta^i{}_j,\quad [\hat{\tilde{p}}^0,\hat{\tilde{q}}_0]=i
\eea
where again we used $C_{ijk}$ is fully symmetric. Especially notice the M2-charges are indeed mutually commuting which with $p^0$ form a complete set of commuting operators. 
The Fourier transform is generated by the quantum version of (\ref{HJ})
\be\label{FO}
\langle\tilde{q}|{q}\rangle=e^{ip^0q_0+2iF(p)}
\ee
where integration over $q_0$ is understood. This establishes a quantum mechanical relation for the wave function in the 4d and {\it formal}  5d charge basis, which allows us to write the topological wavefunction in the 'M2'-charge eigenstates.

Notice also the Hamiltonians depend now only on positions and not on momenta
\be
{\cal H}_i={\tilde q}_0{\tilde q}_i=M^{IJ}{\tilde q}_I{\tilde q}_J~.
\ee
In other words we have found a Jacobi canonical transformation which manifests the conserved charges.

\section{OSV and ${\rm {\bf M}2}$}
In the previous section, we saw that in the Weil-Schr${\rm\ddot o}$dinger representation the topological string holomorphic anomaly equations are {\it formally} identified with membrane charge operators. The present section consists of more speculative discussions on this observation. We hope to put the current ideas on a better footing in future study. 

Given the combination of charges which we formally dubbed the M2's, it's natural to suspect a connection with the Gopakumar-Vafa interpretation of the (A-model) topological string partition function \cite{Gopakumar:1998ii}. However it is not hard to realize one faces seemingly daunting challenges both conceptually and technically to make this vague statement more precise. 

With this goal currently out of reach, here we contend ourselves with some formal investigation which should be considered superficial at best. We will try to recast the OSV relation in a form suited to the M2 charge eigen-states. 

 The OSV conjecture was proposed in \cite{Ooguri:2004zv} as a relation between the topological string partition function and the partition function of the BPS blackholes. Especially it was suggested the topological string amplitudes should capture the subleading corrections to the Bekenstein-Hawking-Wald entropy \cite{LopesCardoso:1998wt}. Starting with \cite{Dabholkar:2004yr}\cite{Vafa:2004qa} it has been tested for various BPS blackholes and in different regimes of charges (see \cite{Pioline:2006ni}).   Following \cite{Verlinde:2004ck} we may write the index for BPS states
\be\label{ind}
\Omega(p,q)=\int d\chi\, e^{-i\chi q}\bra{p-{1\over2}\chi}\hat\Omega\ket{p+{1\over2}\chi}~.
\ee
In the leading approximation $\hat\Omega=\ket{\Psi_{top}}\bra{\Psi_{top}}$ and the index can be considered as the Wigner distribution associated to the topological wavefunciton $\ket{\Psi_{top}}$. To re-write it in terms of the fictitious M-theory charges, we need to transform from $\ket{p}=\ket{p^i,p^0}$ to $\ket{\rm M2}=\ket{\tilde q_i,p^0}$ basis. Applying (\ref{FO}), we find
\bea\label{wigner}
\Omega(p,q)&=&\sum\limits_{\tilde q',\tilde q''}\int d\chi e^{-i\chi q}\bra{p\!-\!\chi/2}{\tilde q'}\rangle\bra{\tilde q'}\hat\Omega\ket{\tilde q''}\langle{\tilde q''}\ket{p\!+\!\chi/2}\nonumber\\
&=&\sum\limits_{\tilde q',\tilde q''}\int d\chi\,e^{-i\chi (q+{\tilde q'+\tilde q''\over2})}e^{2iF(p-{\chi\over2})-2iF(p+{\chi\over2})}\nonumber\\
& &\times\, e^{ip^i(\tilde q_i'-\tilde q_i'')}\,\bra{\tilde q'_i,p^0\!-\!{\chi^0/2}}\hat\Omega\ket{\tilde q''_i,p^0\!+\!{\chi^0/2}}~.\nonumber\\
\eea
Notice we made the index $i$ explicit when the value 0 is not summed over. While we could evaluate this index using saddle point approximation, it will be much simpler to calculate its Legendre transform. The entropy in the M2-charge representation reads
\bea\label{entm}
Z_{bh}(p,\phi)&=&\sum\limits_q\Omega(p,q)e^{\phi\cdot q}\nonumber\\
&=&e^{2iF(p+i{\phi\over2})-2iF(p-i{\phi\over2})}\sum\limits_{\tilde q'+\tilde q'',\atop\tilde q'-\tilde q''}e^{-\phi {\tilde q'+\tilde q''\over2}}\nonumber\\
&&\times e^{ip^i(\tilde q_i'-\tilde q_i'')}\bra{\tilde q'_i,p^0\!+\!i{\phi^0/2}}\hat\Omega\ket{\tilde q''_i,p^0\!-\!i{\phi^0/2}}\nonumber\\
&=&e^{2iF(p+i{\phi\over2})-2iF(p-i{\phi\over2})}\sum\limits_{\tilde q={\tilde q'+\tilde q''\over2},\atop\tilde k={\tilde q'-\tilde q''\over2}}e^{-\tilde q \phi}\,e^{2ip^i\tilde k_i}\nonumber\\
&&\times\bra{\tilde q_i+\tilde k_i,p^0\!+\!i{\phi^0/2}}\hat\Omega\ket{\tilde q_i-\tilde k_i,p^0\!-\!i{\phi^0/2}}~.\nonumber\\
\eea
Now let us try to disentangle this expression a little bit. First we may regard the summation
\be
\tilde\Omega(\tilde q,\tilde p)=\sum\limits_{k}e^{2ip^i k_i}\,\bra{\tilde q_i+ k_i,p^0\!+\!i{\phi^0/2}}\hat\Omega\ket{\tilde q_i-k_i,p^0\!-\!i{\phi^0/2}}
\ee
as a discrete counterpart of the Wigner distribution (\ref{ind}) with however the roles of position and momentum reversed. In addition, the 'nut' charge $p^0$ is singled out and not summed over\footnote{One may consider this as suggesting the different nut charges correspond to super-selection sectors.}. Recall that ${\tilde p}^i=p^i$ and ${\tilde p}^0=-q_0$ we can rewrite (\ref{entm}) as
\be
Z_{bh}(\tilde p,\phi)=\sum\limits_{\tilde q}\,e^{-\phi\cdot\tilde q }e^{-f_0(\tilde p,\phi)}\,\tilde\Omega(\tilde q,\tilde p)
\ee
where $f_0=-2i(F-\bar F)=2{\rm Im}F(p+i{\phi\over2})$ is in fact the genus 0 part of the blackhole partition function. A notable change from (\ref{entm}) is that the electric potential $\phi$ has flipped its sign. 
We find a similarity in this result with that of \cite{Gaiotto:2006ns}. There the factorization of the blackhole partition function was given a physical meaning as arising from antipodal M2 and anti-M2 branes on the horizon. In our case, taking  $\hat\Omega=\ket{\Psi_{top}}\bra{\Psi_{top}}$ we indeed also find the factorization into contributions from M2 and anti-M2 states.

Although we have made some interesting observations, the physical significance of our findings is currently unclear. The holomorphic anomaly equations are by nature perturbative, however it is possible they are relevant in finding new non-perturbative objects in view of the S-duality conjecture for topological string theories \cite{Nekrasov:2004js}, \cite{Kapustin:2004jm}.


{\bf Acknowledgements} We are grateful to Mina Aganagic, Wu-yen Chuang, Sebas de Haro, Tony Pantev and especially Andy Neitzke for insightful discussions.

\end{document}